\documentclass[a4paper,10pt,twoside]{cpc-hepnp}
\usepackage{upgreek,fancyhdr}
\usepackage{multicol}
\usepackage{graphicx}
\usepackage{booktabs}
\usepackage{amssymb,bm,mathrsfs,bbm,amscd}
\usepackage[tbtags]{amsmath}
\usepackage{lastpage}
\usepackage{wrapfig}
\usepackage{lineno}

\begin{document}



\title{Study of BESIII Trigger Efficiencies with the 2018 $J/\psi$ Data}

\maketitle
\begin{small}
\begin{center}
M.~Ablikim$^{1}$, M.~N.~Achasov$^{10,c}$, P.~Adlarson$^{64}$, S. ~Ahmed$^{15}$,
M.~Albrecht$^{4}$, A.~Amoroso$^{63A,63C}$, Q.~An$^{60,48}$, ~Anita$^{21}$, Y.~Bai$^{47}$,
O.~Bakina$^{29}$
% ...
\\
\vspace{0.2cm}
(BESIII Collaboration)\\
\vspace{0.2cm} {\it
$^{1}$ Institute of High Energy Physics, Beijing 100049, People's Republic of China\\
% ...
$^{4}$ Bochum Ruhr-University, D-44780 Bochum, Germany\\
% ...
$^{10}$ G.I. Budker Institute of Nuclear Physics SB RAS (BINP), Novosibirsk 630090, Russia\\
% ...
$^{15}$ Helmholtz Institute Mainz, Johann-Joachim-Becher-Weg 45, D-55099 Mainz, Germany\\
% ...
$^{21}$ Indian Institute of Technology Madras, Chennai 600036, India\\
% ...
$^{29}$ Joint Institute for Nuclear Research, 141980 Dubna, Moscow region, Russia\\
% ...
$^{47}$ Southeast University, Nanjing 211100, People's Republic of China\\
$^{48}$ State Key Laboratory of Particle Detection and Electronics, Beijing 100049, Hefei 230026, People's Republic of China\\
% ...
$^{60}$ University of Science and Technology of China, Hefei 230026, People's Republic of China\\
% ...
$^{63}$ (A)University of Turin, I-10125, Turin, Italy; (B)University of Eastern Piedmont, I-15121, Alessandria, Italy; (C)INFN, I-10125, Turin, Italy\\
$^{64}$ Uppsala University, Box 516, SE-75120 Uppsala, Sweden\\
% ...
$^{c}$ Also at the Novosibirsk State University, Novosibirsk, 630090, Russia\\
% ...
}\end{center}

\vspace{0.4cm}
\end{small}


\begin{abstract}
Using a dedicated data sample taken in 2018 on the $J/\psi$ peak, we perform a detailed study of the trigger efficiencies of the BESIII detector. The efficiencies are determined from three representative physics processes, namely Bhabha-scattering, dimuon production and generic hadronic events with charged particles. The combined efficiency of all active triggers approaches $100\%$ in most cases with uncertainties small enough as not to affect most physics analyses.
\end{abstract}


\begin{multicols}{2}

\section{Introduction}
The Beijing electron-positron collider (BEPCII) is a double-ring multi-bunch $e^{+}e^{-}$ collider with a design luminosity of
$1 \times 10^{33}\; \mathrm{cm^{-2}\, s^{-1}}$,
optimized for a center-of-mass energy of $2\times 1.89$ GeV,
an increase by a factor of 100 over its predecessor.
 The Beijing Spectrometer III (BESIII) detector operating at BEPCII is a multipurpose detector designed for the precision study of $\tau-$charm physics~\cite{Asner:2008nq,bes3:design}.

BEPCII collides electron and positron bunches at a frequency of 125~MHz. The main backgrounds in BESIII are caused by lost beam particles and their interaction with the detector, and the background event rate is estimated to be about 13~MHz~\cite{bes3:detector}. In comparison, the signal rate at the $J/\psi$ resonance is about 2~kHz and the BESIII data acquisition system can record events at a rate of up to 4~kHz. The task of the trigger system is thus to suppress backgrounds by more than three orders of magnitude whilst maintaining a high efficiency for signal events.

Monitoring carefully the trigger efficiency is important in order not to lose events due to inefficient triggers.  A trigger efficiency study was performed in 2010 for data samples of $J/\psi$ and $\psi(2S)$ events recorded in 2009~\cite{bes3:trigger-2009}. Slightly changed trigger conditions after 2012 motivate the study presented here.

\begin{center}
\tabcaption{\label{trigger:con} Trigger conditions.}
\scriptsize
\begin{tabular*}{80mm}{lll}
  \toprule
  No. & Trigger Condition & Comments  \\ \hline\hline
  \multicolumn{3}{c}{Electromagnetic calorimeter (EMC)}\\\hline
   0 & NClus.GE.1 & Number of Clusters $\ge$ 1 \\
   1 & NClus.GE.2 & Number of Clusters $\ge$ 2\\
   2 & BClus$_{-}$BB & Barrel Cluster Back to Back\\
   3 & EClus$_{-}$BB & Endcap Cluster Back to Back\\
   4 & Clus$_{-}$Z &Cluster Balance in $z$ direction\\
   5 & BClus$_{-}$Phi& Barrel Cluster Balance in $\phi$ direction\\
   6 & EClus$_{-}$Phi & Endcap Cluster Balance in $\phi$ direction\\
   7 & BEtot$_{-}$H & Barrel total Energy, Higher threshold\\
   8 & EEtot$_{-}$H & Endcap total Energy, Higher threshold\\
   9 & Etot$_{-}$L & Total Energy, Lower threshold\\
  10 & Etot$_{-}$M & Total Energy, Middle threshold\\
  11 &BL$_{-}$EnZ & Energy Balance in $z$ direction\\
  12 &NBClus.GE.1 & Number of Barrel Clusters $\ge$ 1 \\
  13 &NEClus.GE.1 & Number of Endcap Clusters $\ge$ 1\\
  14 &BL$_{-}$BBLK& Barrel Energy Block Balance\\
  15 &BL$_{-}$EBLK& Endcap Energy Block Balance\\\hline\hline

  \multicolumn{3}{c}{Time of flight system (ToF)}\\\hline
  16 &ETOF$_{-}$BB &Endcap TOF Back to Back\\
  17 &BTOF$_{-}$BB &Barrel TOF Back to Back\\
  18 &NETOF.GE.2 &Number of Endcap TOF hits $\ge$ 2\\
  19 &NETOF.GE.1 &Number of Endcap TOF hits $\ge$ 1\\
  20 &NBTOF.GE.2 &Number of Barrel TOF hits $\ge$ 2\\
  21 &NBTOF.GE.1 &Number of Barrel TOF hits $\ge$ 1\\
  22 &NTOF.GE.1 &Number of TOF hits $\ge$ 1\\\hline\hline

  \multicolumn{3}{c}{Muon counter (MUC)}\\\hline
  32 &NABMU.GE.1&Barrel Tracks number $\ge$ 1 for A\\
  33 &NAEMU.GE.1&Endcap Tracks number $\ge$ 1 for A\\
  34 &NCBMU.GE.1&Barrel Tracks number $\ge$ 1 for C\\
  35 &NCEMU.GE.1&Endcap Tracks number $\ge$ 1 for C\\
  36 &CBMU$_{-}$BB&Barrel Track Back to Back for C\\
  37 &CEMU$_{-}$BB&Endcap Track Back to Back for C\\\hline
\multicolumn{3}{c}{A: 2 of 4 Tracking ; C: 3 of 4 Tracking}\\\hline\hline

   \multicolumn{3}{c}{Main drift chamber (MDC)}\\\hline
  38 &STrk$_{-}$BB &Short Tracks Back to Back\\
  39 &NSTrk.GE.N & Number of Short Tracks $\ge$ N\\
  40 &NSTrk.GE.2 & Number of Short Tracks $\ge$ 2\\
  41 &NSTrk.GE.1 & Number of Short Tracks $\ge$ 1\\
  42 &LTrk$_{-}$BB &Long Tracks Back to Back\\
  43 &NLTrk.GE.N &Number of Long Tracks $\ge$ N\\
  44 &NLTrk.GE.2 &Number of Long Tracks $\ge$ 2\\
  45 &NLTrk.GE.1 &Number of Long Tracks $\ge$ 1\\
  46 &NItrk.GE.2 &Number of Inner Tracks $\ge$ 2\\
  47 &NItrk.GE.1 &Number of Inner Tracks $\ge$ 1\\
\bottomrule
\end{tabular*}
\vspace{0mm}
\end{center}
\vspace{0mm}
The BESIII trigger system combines the information from the electromagnetic calorimeter (EMC), the
main drift chamber (MDC), the time-of-flight system (TOF) and the muon counter (MUC) to form a
total of 48 trigger conditions (Table~\ref{trigger:con}) to select for readout the interesting interactions. A detailed description of the trigger system can be found in Refs.~\cite{bes3:design,bes3:trigger}. The trigger conditions are combined into 16 trigger channels (Table~\ref{trigger:channel}) by the global trigger logic (GTL). The trigger conditions included in trigger channel 12 are delayed by 576~ns in order to distinguish the neutral events from charged events. The event is read out if any enabled trigger channel is active. Compared to earlier data taking periods, for the 2018 $J/\psi$ data taking, CH03 described in Table~\ref{trigger:channel} had to be disabled due to increased noise in the MDC, and CH09 described in Table~\ref{trigger:channel} was added as a high
efficiency selection for neutral events with precise timing information.

\begin{center}
\tabcaption{ \label{trigger:channel} Trigger channels.}
\tiny
\begin{tabular*}{80mm}{lll}
\toprule
Channel & Conditions combination & Comments\\ \hline\hline
CH01 & NEClus.GE.1\&\& NETOF.GE.1\&\& STrk$_{-}$BB & For Charged\\
CH02 & NBClus.GE.1\&\& NBTOF.GE.2\&\& NLtrk.GE.2 & For Charged\\
CH03 & NBTOF.GE.2\&\& NLtrk.GE.2 & Not used\\
CH04 & BTOF$_{-}$BB\&\& LTrk$_{-}$BB & For Charged\\
CH05 & Etot$_{-}$L\&\& NBTOF.GE.1\&\& NLtrk.GE.1 & For Charged\\
CH06 & NBClus.GE.1\&\& NBTOF.GE.1\&\& NLtrk.GE.2 & For Charged\\
CH07 & - & Not used\\
CH08 & - & Not used\\
CH09 & NClus.GE.1\&\& BEtot$_{-}$H & For Neutral\\
CH10 & - & Random\\
CH11 & NBTOF.GE.2\&\& LTrk$_{-}$BB & Not used\\
CH12 & NClus.GE.2\&\& Etot$_{-}$M & Delayed Neutral\\
CH13 & Etot$_{-}$L\&\& NTOF.GE.1 & Not used\\
CH14 & BTOF$_{-}$BB & Not used\\
CH15 & NClus.GE.1 & Not used\\
CH16 & ECLUS$_{-}$BB & Not used\\
\bottomrule
\end{tabular*}
\vspace{0mm}
\end{center}

Using a similar approach as described in Ref.~\cite{bes3:trigger-2009}, we study the trigger efficiency for the $J/\psi$ events taken in 2018 in order to understand the performance for the updated trigger system.

\section{Data Set}
\subsection{Trigger menu for the 2018 data taking}
Table~\ref{trigger:2018jpsi} shows the trigger menu used for the 2018 $J/\psi$ data taking campaign, which has not changed since 2012 with the exception of CH03 mentioned above. The enabled channels are categorized into three almost independent groups, namely endcap charged, barrel charged and neutral.

\begin{center}
\tabcaption{ \label{trigger:2018jpsi} Trigger menu for 2018 $J/\psi$ data taking.}
\tiny
\begin{tabular}{ll|l}
\toprule
Channel & Conditions & Group\\ \hline\hline
CH01 & NEClus.GE.1\&\& NETOF.GE.1\&\& STrk$_{-}$BB & Endcap Charged\\\hline
CH02 & NBClus.GE.1\&\& NBTOF.GE.2\&\& NLtrk.GE.2 &\\
CH04 & BTOF$_{-}$BB\&\& LTrk$_{-}$BB & Barrel Charged\\
CH05 & Etot$_{-}$L\&\& NBTOF.GE.1\&\& NLtrk.GE.1& \\
CH06 & NBClus.GE.1\&\& NBTOF.GE.1\&\& NLtrk.GE.2 &\\\hline
CH09 & NClus.GE.1\&\& BEtot$_{-}$H & Neutral\\
CH12 & NClus.GE.2\&\& Etot$_{-}$M &\\
\bottomrule
\end{tabular}
\vspace{0mm}
\end{center}

\subsection{Data sample for trigger study}
To study the trigger efficiency, we took two dedicated runs (run 56199 and run 56200)
where a single trigger was enabled in order to determine the efficiencies of all trigger conditions using a set of independent conditions. The corresponding trigger menus are shown in Table~\ref{trigger:2018jpsitest}.

\begin{center}
\tabcaption{\label{trigger:2018jpsitest} Trigger menu for the 2018 $J/\psi$ test runs.}
\footnotesize
\begin{tabular}{l|l}
\toprule
Channel & Run number\\ \hline
CH03 & 56199\\\hline
CH12 & 56200\\
\bottomrule
\end{tabular}
\vspace{0mm}
\end{center}

\section{Control Sample Selection}
As widely used in BESIII physics analyses, only the tracks with a polar angle $\theta$ (defined relative to the positron beam direction) for which $|\cos\theta| \leq 0.93$ are taken into account. The barrel region is defined as $|\cos\theta|<0.8$, and the endcap region as $0.86<|\cos\theta|<0.92$.  The definitions of ``barrel" and ``endcap" are slightly between the analysis definitions and the trigger system, for which the ``barrel" and ``endcap" are decided by the structure of the sub-detector (such as MDC, EMC,...).  The charged lepton or hadron selection defines good charged particle tracks as those with a distance of closest approach to the
interaction point within 10 cm along the beam direction and 1 cm in the plane transverse to the beam direction.
The control samples are selected similarly to those in Ref.~\cite{bes3:trigger-2009} and are described in the following subsections.

\subsection{Bhabha Event Selection}

 To select Bhabha events, two EMC clusters are required to have an opening angle larger than $166^{\circ}$ and an energy difference within $10\%$ of the center-of-mass energy:
  $$\frac{|E_{\rm emc}(e^{+})+E_{\rm emc}(e^{-})-3.097|}{3.097}\leq 10\% ~.$$
Two oppositely  charged good tracks in the MDC with an opening angle of more than $175^{\circ}$ are selected.
Potential backgrounds have been investigated using an inclusive Monte Carlo (MC) samples, which consists of the production of the $J/\psi$ resonance, and the continuum processes incorporated in $KKMC$~\cite{bes3:kkmc}, where the known decay modes are modeled with $EVTGEN$~\cite{bes3:evtgen} using branching fractions taken from the Particle Data Group~\cite{PDG}, and the remaining unknown decays from the charmonium states are generated with $LUNDCHARM$~\cite{lundcharm}. Using this sample, the impurity of the selected Bhabha sample is determined to be about $1.6\times 10^{-6}$.

\subsection{Dimuon Event Selection}
  To select dimuon candidate events,  two oppositely charged good tracks are required to have an opening angle of at least $178^{\circ}$. In addition, we require that the momentum of each track is less than 2~GeV/$c$, and that the deposited energy in the EMC is less than 0.7~GeV. The total four-momentum $(E/c, P_{x}, P_{y}, P_{z})$ is required to fall into the range (2.8 to 3.3, -0.1 to 0.1, -0.1 to 0.1, -0.2 to 0.2) GeV/$c$, assuming that both tracks are muons.
By using the inclusive $J/\psi$ decay MC sample, we investigate potential backgrounds, and find the background levels to be less than 0.4\%.

\subsection{Charged Hadronic Event Selection}
For the hadron selection, two or more good tracks are required in the MDC. If there are exactly two tracks, the opening angle between them is required to be less than $170^{\circ}$ in order to suppress Bhabha and dimuon backgrounds.

\section{Trigger Efficiency Determination}
All of the 2018 $J/\psi$ data (runs 53207--56520) available were taken using the same trigger conditions,  and the main challenge in the efficiency determination is to reduce any bias to a minimum. Thus we use the two test runs triggered by independent trigger channels (Table ~\ref{trigger:2018jpsitest}) to determine the trigger efficiencies.

\subsection{Determination of Trigger Efficiencies}
The trigger efficiency for each trigger condition/trigger channel ($\varepsilon_\text{cond./ch}$) can be calculated using
$$\varepsilon_\text{cond./ch}=\frac{N_\text{(sel,~trig.condition/channel)}}{N_\text{sel}} ~,$$
where``$N$'' stands for the number of events, the label ``sel'' for events passing the physics selection, and ``trig.condition/channel'' for events in which the trigger condition/channel under study is active. The efficiencies of the trigger conditions which have been used for the 2018 $J/\psi$ data taking are listed in Table~\ref{trigger:coneff}. The Clopper-Pearson method~\cite{error} has been used to estimate the confidence interval at the confidence level of $1-\alpha = 0.6827 (1\sigma)$.
It should be noted that the number of prongs for hadronic events refers to the number of charged tracks in the full detector, not only in the barrel or endcap.

\subsection{Determination of Trigger Channel Efficiencies}
The efficiency of the trigger channels can be determined similar to the efficiency of the trigger conditions if a fully independent trigger channel exists. Otherwise, a mathematical combination of the condition efficiencies has to be performed.  By considering the three almost independent groups of channels shown in Table~\ref{trigger:2018jpsi}, we can obtain the trigger channel efficiencies for 2018 $J/\psi$ data taking as follows:
\end{multicols}
\vspace{0.5cm}
\ruleup
$$\varepsilon_\text{final} = g_{1}+g_{2}+g_{3}-(g_{1}g_{2}+g_{1}g_{3}+g_{2}g_{3})+ g_{1}g_{2}g_{3} $$
\noindent  where $g_{n}$ is the efficiency of the $n^{th}$ group of trigger channels.
\begin{center}
\tabcaption{\label{trigger:coneff} Trigger condition efficiencies (in $\%$) (Note: The relative uncertainties of the items with no uncertainties indicated are less than $0.01\%$).}
\footnotesize
\newcommand\ST{\rule[-0.6em]{0pt}{1.8em}}
\begin{tabular}{|c|r|r|r|r|r|r|r|r|}
\toprule
  &  &  & \multicolumn{2}{|c|}{Bhabha} & \multicolumn{2}{|c|}{Dimuon} &2-prong & 4-prong\\\cline{1-7}
  & GTL & Condition &Barrel & Endcap &Barrel &Endcap & & \\\hline

   & 0 & NClus.GE.1
   & 100.00 & 100.00$^{+0.00}_{-0.41}$
   & 99.93$\pm$0.01 & 94.74$^{+4.35}_{-11.09}$
   & 99.64$\pm$0.01
   & 99.97 \ST \\\cline{2-9}

   & 1 & NClus.GE.2
   &98.69$\pm$0.03 & 98.20$^{+0.62}_{-0.87}$
   &95.14$\pm$0.08 & 84.21$^{+8.47}_{-13.01}$
   &98.01$^{+0.03}_{-0.02}$
   &99.63$^{+0.01}_{-0.02}$ \ST \\\cline{2-9}

  E&7 & BEtot$_{-}$H
   &100.00 &0.17$\pm$0.02
   &0.68$\pm$0.03 &4.81$^{+2.06}_{-3.12}$
   &89.88$\pm$0.04
   &93.25 $^{+0.03}_{-0.04}$ \ST \\\cline{2-9}

  M&9 & Etot$_{-}$L
   &100.00 &100.00$^{+0.00}_{-0.41}$
   & 99.82$\pm$0.01 &100.00$^{+0.00}_{-9.24}$
   &99.63$\pm$0.01
   &99.99 \ST \\\cline{2-9}

  C&10 & Etot$_{-}$M
   &100.00 &100.00$^{+0.00}_{-0.41}$
   &10.25$\pm$0.11 & 0.00$^{+0.09}_{-0.00}$
   &97.01$\pm$0.03
   &99.44$\pm$0.02 \ST  \\\cline{2-9}

   &12 &NBClus.GE.1
   &100.00 &0.99$\pm$0.01
   &99.93$\pm$0.01 &0.00$^{+0.09}_{-0.00}$
   &99.34$\pm$0.01
   &99.90$\pm$0.01 \ST \\\cline{2-9}

   &13 &NEClus.GE.1
   &0.94$\pm$0.02        &100.00$^{+0.00}_{-0.41}$
   &1.68$^{+0.04}_{-0.05}$ &94.74$^{+4.35}_{-11.09}$
   &36.93$\pm$0.06
   &41.85$\pm$0.07 \ST \\\hline

   &17 &BTOF$_{-}$BB
   &98.81$\pm$0.01 &0.62$^{+0.02}_{-0.03}$
   &99.98$\pm$0.01 &0.00$^{+0.02}_{-0.00}$
   &57.21$\pm$0.06
   &83.21$\pm$0.05 \ST \\\cline{2-9}

  T&19 &NETOF.GE.1
   &61.98$\pm$0.09 & 99.90$^{+0.00}_{-0.01}$
   &60.08$\pm$0.17 & 100.00$^{+0.00}_{-2.14}$
   &74.69$^{+0.05}_{-0.06}$
   &77.87$\pm$0.06 \ST \\\cline{2-9}

  O&20 &NBTOF.GE.2
   &99.69$^{+0.01}_{-0.02}$ &3.69$\pm$0.06
   &99.89$^{+0.04}_{-0.06}$ &7.06$^{+2.76}_{-3.99}$
   &87.81$^{+0.05}_{-0.06}$
   &99.04$\pm$0.02 \ST \\\cline{2-9}

  F&21 &NBTOF.GE.1
   &100.00 &41.89$\pm$0.14
   &100.00 &36.47$^{+5.60}_{-5.95}$
   &99.63$\pm$0.01
   &99.96 \ST \\\hline

   &38 &STrk$_{-}$BB
   &99.93$^{+0.00}_{-0.01}$  &99.95$\pm$0.01
   &99.95$\pm$0.01  &100.00$^{+0.00}_{-1.75}$
   &46.62$\pm$0.06
   &83.01$^{+0.05}_{-0.06}$ \ST \\\cline{2-9}

  M&42 &LTrk$_{-}$BB
   &99.91$^{+0.00}_{-0.01}$ &6.96$^{+0.07}_{-0.08}$
   &99.95$^{+0.01}_{-0.02}$  &11.54$^{+4.03}_{-3.19}$
   &37.34$\pm$0.06
   &76.21$\pm$0.06 \ST \\\cline{2-9}

  D&44 &NLTrk.GE.2
   &99.90$^{+0.00}_{-0.01} $ &21.74$\pm$0.12
   &99.87$^{+0.05}_{-0.06}$  &18.82$^{+5.22}_{-4.39}$
   &93.68$\pm$0.05
   &99.86$\pm$0.02 \ST \\\cline{2-9}

  C&45 &NLTrk.GE.1
   &100.00  &38.92$^{+0.13}_{-0.14}$
   &100.00  &30.59$^{+5.80}_{-5.30}$
   &99.67$\pm$0.01
   &99.98 \ST \\
\bottomrule
\end{tabular}
\vspace{0mm}
\end{center}
\vspace{0mm}

\begin{multicols}{2}
The logical relationship between trigger channels (Table~\ref{trigger:2018jpsi}) is ``or'', and in each trigger channel, the relationship between trigger conditions is ``and'', so the efficiencies for the groups of trigger channels are the sum of all efficiencies of the channels in question with the overlap of the channels subtracted. The efficiencies of the groups of trigger channels can be calculated as:
\begin{eqnarray*}
  g_{1} &=& c_1 \\
  g_{2} &=& A-B+C-D \\
  g_{3} &=& E-F
\end{eqnarray*}
\noindent and,
\end{multicols}
\vspace{0.5cm}
\ruleup
\begin{eqnarray*}
  A &=& c_2+c_4+c_5+c_6 \\
  B &=& c_2c_4+c_2c_5+c_2c_6+c_4c_5+c_4c_6+c_5c_6 \\
    &=& c_2\cdot P(4|2)+c_2\cdot P(5|2)+c_2\cdot P(6|2)+c_4\cdot P(5|4)+c_6\cdot P(4|6)+c_6\cdot P(5|6) \\
  C &=& c_2c_4c_5+c_2c_4c_6+c_2c_5c_6+c_4c_5c_6 \\
    &=& c_2\cdot P(4,5|2)+c_2\cdot P(4,6|2)+c_2\cdot P(5,6|2)+c_6\cdot P(4,5|6)\\
  D &=& c_2c_4c_5c_6 \\
    &=& c_2\cdot P(4,5,6|2)\\
  E &=& c_9+c_{12} \\
  F &=& c_9c_{12}\\
    &=& c_9\cdot P(12|9)
\end{eqnarray*}
\ruledown
\vspace{0.5cm}
\begin{multicols}{2}
\noindent  where $A$ and $E$ are the sum of trigger channel efficiencies in the group. $B$, $D$ and $F$ are the overlap efficiencies for double-counting parts in $A$ and $E$.  $C$ is the efficiency double-counted in $B$ and $D$. $cn$ is the efficiency of $n^{th}$ channel. $P(n,\ldots|m)$ is a conditional probability, $i.e.$ how many events of condition $(n,\ldots)$ involved into the condition $m$,  which is the overlap/correlations if the trigger channels are not independent of each other in the same group.

Using the combination methods outlined above, the overall efficiencies of the trigger channels and global trigger efficiencies are given in Table~\ref{trigger:cheff}.
\end{multicols}

\begin{table}
\begin{center}
\caption{\label{trigger:cheff} Global trigger efficiencies (in $\%$) (Note: The relative uncertainties of the items with no uncertainties given are less than $0.01\%$)}
\footnotesize
\newcommand\ST{\rule[-0.6em]{0pt}{1.8em}}
\begin{tabular}{|c|r|r|r|r|r|r|}
\toprule
  &   \multicolumn{2}{|c|}{Bhabha} & \multicolumn{2}{|c|}{Dimuon} &2-prong &4-prong\\\cline{2-7}

   Channel &Barrel & Endcap &Barrel & Endcap & & \\\hline

  CH01
   &0.65$\pm$0.02 & 99.10$^{+0.43}_{-0.70}$
   &0.63$\pm$0.03 & 99.04$^{+0.96}_{-11.09}$
   &15.88$\pm$0.04
   &31.30$^{+0.03}_{-0.05}$ \ST \\\hline

  CH02
   &99.60$\pm$ 0.02 & 0.03$\pm$0.01
   &99.76$^{+0.06}_{-0.08}$ &1.18$^{+0.85}_{-0.78}$
   &84.88$\pm$0.06
   &98.97$\pm$0.02 \ST \\\hline

  CH04
   &99.73$\pm$ 0.01 & 0.06$\pm$0.01
   &99.92$\pm$0.01  &0.00$^{+0.02}_{-0.00}$
   &29.15$\pm$0.05
   &67.36$\pm$0.07 \ST \\\hline

  CH05
   &100.00 &17.45$\pm$0.11
   &99.82$\pm$0.01  &9.41$^{+2.32}_{-1.69}$
   &99.04$\pm$0.01
   &99.94 \ST \\\hline

  CH06
   &99.90$\pm$0.01 & 0.15$^{+0.01}_{-0.02}$
   &99.87$^{+0.04}_{-0.06}$ &2.35$^{+1.02}_{-0.72}$\
   &93.22$^{+0.05}_{-0.06}$
   &99.78$\pm$0.01 \ST \\\hline

  CH09
   &100.00 & 0.17$\pm$0.01
   &0.68$\pm$0.03   &5.88$^{+2.79}_{-1.52}$
   &89.85$\pm$0.04
   &93.23$\pm$0.04 \ST \\\hline

 CH12
   &98.69$\pm$0.03 & 98.20$^{+0.62}_{-0.87}$
   &9.79$\pm$0.12  & 0.00$^{+0.09}_{-0.00}$
   &96.42$^{+0.04}_{-0.03}$
   &99.22$\pm$0.02 \ST \\\hline\hline

 Barrel Charged
   &100.00$^{+0.00}_{-0.02}$ & 17.45$^{+6.61}_{-6.91}$
   &99.95$^{+0.05}_{-0.10}$  & 9.41$^{+8.25}_{-7.06}$
   &99.04$\pm$0.19
   &99.94$^{+0.06}_{-0.11}$ \ST \\\hline

 Endcap Charged
   &0.65$\pm$0.02 & 99.10$^{+0.43}_{-0.70}$
   &0.63$\pm$0.03 & 99.04$^{+0.96}_{-11.09}$
   &15.88$\pm$0.04
   &31.30$^{+0.03}_{-0.05}$ \ST \\\hline

 Neutral
   &100.00$^{+0.00}_{-0.03}$ & 98.20$^{+1.80}_{-5.84}$
   &9.81$\pm$0.45  &5.88$^{+2.79}_{-1.52}$
   &96.71$^{+0.06}_{-0.05}$
   &99.32$\pm$0.05 \ST \\\hline\hline

 Total
   &100.00 & 99.99$^{+0.01}_{-0.04}$
   &99.96$^{+0.04}_{-0.09}$ & 99.33$^{+0.67}_{-9.46}$
   &99.97$\pm$0.01
   &100.00$^{+0.00}_{-0.01}$ \ST \\
\bottomrule
\end{tabular}
\vspace{0mm}
\end{center}
\end{table}

\begin{multicols}{2}
\section{Summary}
The BESIII trigger system is a fundamental tool for a successful collection of data for physics analyses. With a dedicated data sample collected at the $J/\psi$ peak, the trigger efficiencies for
various physics channels were determined, and found to be close to $100\%$ for most physics cases with small uncertainties. This conclusion is similar to that arrived by the trigger study for the 2009 run~\cite{bes3:trigger-2009}, showing that
there was no significant degradation in almost a decade of running.
As the trigger menu studied here has been used for all data taking since 2012, the results of this study apply to all respective data samples.
For most physics channels, the efficiency of the full trigger menu approaches $100\%$ and can be neglected in physics analyses.
\\

\acknowledgments{The BESIII collaboration thanks the staff of BEPCII and the IHEP computing
center for their strong support. This work is supported in part by National Key Basic
Research Program of China under Contract No. 2015CB856700; National Natural Science
Foundation of China (NSFC) under Contracts Nos. 11625523, 11635010, 11735014, 11822506,
11835012, 11935015, 11935016, 11935018, 11961141012; the Chinese Academy of Sciences (CAS)
Large-Scale Scientific Facility Program; Joint Large-Scale Scientific Facility Funds of the
NSFC and CAS under Contracts Nos. U1732263, U1832207; CAS Key Research Program of Frontier
Sciences under Contracts Nos. QYZDJ-SSW-SLH003, QYZDJ-SSW-SLH040; 100 Talents Program of
CAS; INPAC and Shanghai Key Laboratory for Particle Physics and Cosmology; ERC under
Contract No. 758462; German Research Foundation DFG under Contracts Nos. Collaborative
Research Center CRC 1044, FOR 2359; Istituto Nazionale di Fisica Nucleare, Italy; Ministry
of Development of Turkey under Contract No. DPT2006K-120470; National Science and Technology
fund; Olle Engkvist Foundation under Contract No. 200-0605; STFC (United Kingdom); The Knut and Alice Wallenberg Foundation (Sweden) under
Contract No. 2016.0157; The Royal Society, UK under Contracts Nos. DH140054, DH160214; The
Swedish Research Council; U. S. Department of Energy under Contracts Nos. DE-FG02-05ER41374,
DE-SC-0012069}

\end{multicols}

\end{document}